\documentclass{article}

\usepackage{arxiv}

\usepackage[utf8]{inputenc} 
\usepackage[T1]{fontenc}    
\usepackage{hyperref}       
\usepackage{url}            
\usepackage{booktabs}       
\usepackage{amsfonts}       
\usepackage{nicefrac}       
\usepackage{microtype}      
\usepackage{lipsum}		
\usepackage{graphicx}
\usepackage{natbib}
\usepackage{doi}

\usepackage{graphics} 
\usepackage{epsfig} 
\usepackage{mathptmx} 
\usepackage{times} 
\usepackage{amsmath} 
\usepackage{amssymb}  
\usepackage{hyperref}
\usepackage{algorithm}
\usepackage{algpseudocode}
\usepackage{algorithmicx}
\usepackage{multicol}

\DeclareMathOperator*{\argmin}{arg\,min}

\title{Minimal Order Recovery through Rank-adaptive Identification}

\author{ {Frédéric Zheng} \\
	School of EECS, KTH\\
	Stockholm, Sweden\\
	\texttt{fzheng@kth.se} 
	\And
	{Yassir Jedra} \\
	LIDS, Dept. of EECS, MIT\\
	Cambridge, MA, USA\\
	\texttt{jedra@mit.edu} 
    \And
	{Alexandre Proutière} \\
	School of EECS, KTH\\
	Stockholm, Sweden\\
	\texttt{alepro@kth.se} 
}

\date{}

\renewcommand{\headeright}{}
\renewcommand{\undertitle}{}
\renewcommand{\shorttitle}{}

\newtheorem{theorem}{Theorem}[section]
\newtheorem{proposition}[theorem]{Proposition}
\newtheorem{lemma}[theorem]{Lemma}

\newtheorem{assumption}[theorem]{Assumption}

\begin{document}

\maketitle

\begin{abstract}
This paper addresses the problem of identifying linear systems from noisy input-output trajectories. We introduce Thresholded Ho-Kalman, an algorithm that leverages a rank-adaptive procedure to estimate a Hankel-like matrix associated with the system. This approach optimally balances the trade-off between accurately inferring key singular values and minimizing approximation errors for the rest. We establish finite-sample Frobenius norm error bounds for the estimated Hankel matrix. Our algorithm further recovers both the system order and its Markov parameters, and we provide upper bounds for the sample complexity required to identify the system order and finite-time error bounds for estimating the Markov parameters. Interestingly, these bounds match those achieved by state-of-the-art algorithms that assume prior knowledge of the system order.
\end{abstract}

\section{Introduction}

We revisit the problem of identifying a linear time-invariant (LTI) system from noisy input-output trajectories. We assume these trajectories are the only available data and, in particular, that the learner has no prior knowledge of the system’s state or even its order. A standard approach to LTI system identification involves first estimating a Hankel-like matrix associated with the system, followed by applying a subspace identification method (SIM) such as the celebrated Ho-Kalman algorithm \cite{kalman1966effective}. Over the past fifty years, SIMs have received considerable attention (see, e.g., \cite{viberg1995subspace,Qin2006,deveen2013} for surveys), but their statistical properties have been studied primarily in the asymptotic regime \cite{Jansson1998,Jansson2000,Bauer2005,Chiuso2004,chiuso2005consistency}. More recently, efforts have been made to analyze the finite-time properties of these methods \cite{oymak2019non,sun2022finite,sarkar2021finite, HeZRH24}, though always under the assumption that the system order is known.

In this work, we aim to develop an identification algorithm that enjoys finite-time statistical guarantees without requiring prior knowledge of the system order. Specifically, we seek to establish upper bounds on the number of samples needed to estimate the Hankel matrix, the system order, and its Markov parameters with a given level of certainty and accuracy. 

We introduce {\it Thresholded Ho-Kalman}, an algorithm that enhances an initial estimate of the Hankel matrix using a universal singular value thresholding procedure. This initial estimate is typically obtained through a least-squares method. The thresholding procedure is designed using novel insights from matrix denoising theory. Our approach adapts to the available data and the noise level of the initial estimator, enabling it to learn an optimal subset of the Hankel singular values. This subset balances the trade-off between estimating its singular values and disregarding the rest. We provide finite-time guarantees for Thresholded Ho-Kalman when learning a system from either a single or multiple trajectories. Specifically, we establish upper bounds on the sample complexity required to identify the system order with a given confidence level. Additionally, we derive finite-time error bounds for the estimation of the Hankel matrix and Markov parameters. Remarkably, these bounds align with those of state-of-the-art algorithms that assume prior knowledge of the system order. We confirm this observation through numerical experiments.

\section{Related work}

Finite-time analysis for system identification has recently seen a surge in interest. Significant advancements have been made recently for fully observable dynamics, see e.g. \cite{simchowitz2018learning,Tsiamis2019FiniteSA,jedra2019sample,jedra2020finite,Lale2021FinitetimeSI,jedra2022finite}. These methods utilize a single, potentially noisy trajectory and sidestep the challenge of determining the system order $n$.

In the more general case of partially observed systems, SIMs are commonly used to recover both $n$ and the Markov parameters (up to a similarity transformation) from the Hankel matrix. When dealing with noisy data, the Hankel matrix must be estimated, either by first estimating the finite impulse response \cite{fazel2013hankel, oymak2019non, sun2022finite} or directly \cite{sarkar2021finite}. These papers provide a finite-time error analysis of the estimator of the Hankel matrix. However, since the rank of the estimated Hankel matrix may not equal $n$, SIMs cannot be directly applied to the estimator. When $n$ is known, \cite{oymak2019non} explores a variant of the Ho-Kalman algorithm to estimate the Markov parameters, providing upper bounds on the identification error. To our knowledge, there is no finite-time analysis for procedures that recover $n$ when it is initially unknown. Existing literature offers only asymptotic guarantees for order estimation \cite{shibata1976selection, moonen1989and, bauer2001order}.








\section{Preliminaries}

\subsection{Notation} We denote by $\Vert . \Vert_2, \Vert.\Vert_F$ operator and Frobenius for matrices and by $\Vert.\Vert_{l_2}$ the Euclidean norm for vectors. Let $x, y \in \mathbb{R}^{n}$ and denote their scalar product by $\langle x, y \rangle$. Let $M\in\mathbb{R}^{m\times n}$ a matrix with SVD $M=\sum_{i=1}^{\min(m,n)}s_i(M)u_iv_i^\top$ where $s_i(M)$ is the $i$-th largest singular value and $u_i$ and $v_i$ are the corresponding singular vectors. For $k\le \min(m,n)$, we denote by $\Pi_k(M)=\sum_{i=1}^ks_i(M)u_iv_i^\top$ the best rank-$k$ approximation of $M$. For any $\xi\ge 0$, we define $M(\xi)=\sum_{i=1}^{\min(m,n)}1_{(s_i(M)\geq \xi)}s_i(M)u_iv_i^\top$. Let $M^\dagger$ the Moore-Penrose inverse of $M$. Finally, the inequality $x\lesssim y$ means that $x$ is smaller than $y$ up to a universal multiplicative constant.





\subsection{Model and Assumptions}
Consider the following discrete-time LTI system:
\begin{align}
\begin{split}
    x_{t+1} &= A x_t + B u_t  \\
    y_{t} &= C x_{t} + z_t  
\end{split}
\end{align}
where $x_t\in \mathbb{R}^n$, $u_t\in \mathbb{R}^{d_u}$, and $y_t\in \mathbb{R}^{d_y}$ denote the state, the input, and the observation at time $t$, respectively. The matrices $A\in \mathbb{R}^{n\times n}, B\in\mathbb{R}^{n\times d_u}, C\in \mathbb{R}^{d_y\times n}$ are assumed to be such that $(A,B)$ is controllable and $(A,C)$ is observable. To simplify the notation, we assume that $x_1=0$. We further assume that the observation noise and the inputs are independent sequences of i.i.d. Gaussian vectors, i.e., $z_t\sim \mathcal{N}(0,\sigma_z^2I_{d_y})$ and $u_t\sim \mathcal{N}(0,\sigma_u^2I_{d_u})$. All our results can be extended to isotropic subGaussian vectors with variance proxy $(\sigma_u,\sigma_z)$. 

Our objective is to estimate the order $n$ of the system and the Markov parameters $A$, $B$, and $C$, from input-output observations. To this aim, we plan to estimate Hankel-like matrices related to the system. For a given window length $\tau\geq 1$, define $G_\tau:=(CB\ldots CA^{2\tau-2}B)\in\mathbb{R}^{d_y\times (2\tau-1)d_u}$. Observe that for $t\geq 2\tau$, this truncated impulse response naturally maps $y_{t}$ to the past $2\tau-1$ inputs $(u_{t-1},...,u_{t-2\tau+1})$ through 

\begin{equation}
    y_t = G_\tau \begin{pmatrix}
    u_{t-1}^\top & 
    \cdots & 
    u_{t-2\tau+1}^\top 
\end{pmatrix}^\top +CA^{2\tau-1}x_{t-2\tau+1} + z_t.
\label{eq:G}
\end{equation}

Define the finite Hankel matrix as

$$H_{\tau}=\begin{pmatrix} CB & \cdots & CA^{\tau-1}B \\
\vdots & \ddots & \vdots \\
CA^{\tau-1}B & \cdots & CA^{2\tau-2}B
\end{pmatrix}\in\mathbb{R}^{\tau d_y\times \tau d_u}.$$
Throughout this paper, we will also make use of the linear map $\mathcal{H}:G_\tau \mapsto H_\tau$.

\medskip\begin{assumption}\label{ass:rank}
    $\tau\geq n+1$.
\end{assumption}
When this assumption holds, it is a well-known result that $H_\tau$ has rank equal to $n$. 



\subsection{Ho-Kalman algorithm and its thresholded version}

When the Hankel matrix $H_\tau$ is known, under Assumption \ref{ass:rank}, one may retrieve first the order $n$ of the system by computing its rank, and then the Markov parameters. To this aim, we can use one of the equivalent versions of the Ho-Kalman factorization algorithm \cite{kalman1966effective}. Let $H_\tau^\rightarrow$ (resp. $H_\tau^{\leftarrow}$) denote the matrices obtained from $H_\tau$ by removing the last (resp. first) column block. When $\tau\geq n+1$, both matrices have a rank equal to $n$. The Ho-Kalman algorithm computes the SVD $USV^\top$ of $H_\tau^\rightarrow$, and extract the matrices $O = U S^{1/2}$ and $Q=S^{1/2} V^\top$. The Markov parameters are finally retrieved: $\bar A= O^\dagger H_\tau^{\leftarrow}  Q^\dagger$, $\bar B$ consists of the first $d_u$ columns of $Q$, and $\bar C$ is the first $d_y$ rows of $O$. These matrices are equal to the true Markov parameters up to an invertible matrix: there exists $P\in \mathbb{R}^{n\times n}$ invertible such that $\bar A=P^{-1}AP$, $\bar{B}=P^{-1}B$ and $\bar{C}=CP$.














To estimate the order of the system and its Markov parameters from input-output samples, we employ a two-step procedure. 
\begin{itemize}
\item[(i)] We begin by obtaining an initial estimator, $\hat{H}_\tau$, of the matrix ${H}_\tau$ via a (possibly regularized) Least-Squares method. This initial estimator $\hat{H}_\tau$ is naturally affected by noise and is likely to be full rank (i.e., of rank $\min(\tau d_y,\tau d_u)> n$). 
\item[(ii)] In the second step, we derive a low-rank approximation,  $\hat{H}_\tau(\xi)$ of $\hat{H}_\tau$ obtained by keeping the singular subspaces associated to singular values above an appropriately chosen threshold $\xi$. The rank of the resulting matrix provides an estimator of the order of the system, and the Markov parameters will be estimated by applying the Ho-Kalman algorithm to $\hat{H}_\tau(\xi)$. The pseudo-code of the second step is detailed in Algorithm \ref{algo:1}.  
\end{itemize}

\begin{algorithm}[H]
\caption{Thresholded Ho-Kalman}
\label{algo:1}
\begin{algorithmic}

\State \textbf{Input} Estimator $\hat H_\tau$, threshold $\xi$


\State$  \hat U,\hat S, \hat V \leftarrow \textrm{SVD of } (\hat H_\tau(\xi))^\rightarrow$ 
\State $\hat O \leftarrow \hat U \hat S^{1/2}$
\State $ \hat Q \leftarrow  \hat S^{1/2} \hat V^\top$
\State $\hat C \leftarrow \text{first $d_y$ rows of }  \hat O$ 
\State $\hat B \leftarrow \text{first $d_u$ columns of } \hat Q$
\State $\hat A \leftarrow  \hat O^\dagger (\hat H_\tau(\xi))^{\leftarrow}  \hat Q^\dagger$

\State \textbf{return} $\textrm{rank}(\hat H_\tau(\xi))$, $\hat A$, $\hat B$, $\hat C$.
\end{algorithmic}
\end{algorithm}

\section{Universal singular value thresholding}

In this section, we analyze the singular value thresholding used in the second step of our estimation procedure and applied to an initial estimator $\hat{H}_\tau$ of the Hankel matrix $H_\tau$. We denote by $Z = \hat{H}_\tau - H_\tau$ the estimation error. We show that, should we have access to a high-probability upper bound on $2\|Z\|_2$, using this bound as the threshold $\xi$ results in a refined estimator $\hat{H}_\tau(\xi)$ with strong performance guarantees. In the following, we  study its Frobenius error $\| \hat{H}_\tau(\xi) - H_\tau\|_F$, and its rank. 

\medskip
\begin{proposition} For any $k=0,\ldots,n$, we have: 
$$\Vert \Pi_k(\hat{H}_\tau)-H_\tau\Vert_F^2\leq 18(k\Vert Z\Vert_2^2+ \sum_{i=k+1}^ns_i^2(H_\tau)).$$
Furthermore for all $\xi \ge 2\Vert Z\Vert= 2\| \hat{H}_\tau -H_\tau\|_2$, we have:
    \begin{equation}\label{eq:rankadapt}
    \Vert \hat{H}_\tau(\xi) -H_\tau\Vert_F^2\leq 18\min_{k=0,...n}\left( 4 k\xi^2+ \sum_{i=k+1}^{n}s_i^2(H_\tau)\right).
\end{equation}
\label{prop:threshold}
\end{proposition}

It is worth comparing the result of this proposition to that derived in  \cite{chatterjee2015matrix}. There, Chatterjee proposes a threshold of the form $\xi =(1+u)\Vert Z\Vert_2$, for $u>0$, and establishes that 
$
\|\hat{H}_\tau(\xi)-H_\tau\Vert_F^2\leq f(u) \Vert Z\Vert_2 \Vert H_\tau\Vert_1
$
where $f(u) = ((4+2u)\sqrt{\frac{2}{u}}+\sqrt{2+u})^2$, and $\Vert . \Vert_1$ is the nuclear norm. This upper bound is conservative as it depends on the nuclear norm of $H_\tau$, and consequently on all the singular values of $H_\tau$. In contrast, our upper bound only involves the singular values smaller than $\xi$.  

Our upper bound (\ref{eq:rankadapt}) can be interpreted as follows. The Frobenius error of the thresholded estimator can be decomposed into the cost $k\xi^2$ of estimating $k$ top singular values of $H_\tau$ and the cost $\sum_{i=k+1}^{n}s_i^2(H_\tau)$ of not identifying the remaining singular values. (\ref{eq:rankadapt}) means that applying a singular value thresholding at $\xi$ actually optimizes the trade-off (the 'min' over $k$) between these two costs, and defines an {\it effective rank} $k_\xi := \mathrm{rank}(\hat H_\tau(\xi))$. This effective rank corresponds to the low-rank approximation of $H_\tau$ that should be estimated. The threshold $\xi$ needs to depend on $Z$, and hence on the number of available input-output samples. In Sections \ref{sec:single} and \ref{sec:multi}, we show how to select the threshold in an adaptive manner.

The next lemma is a consequence of Weyl’s inequality and provides basic properties of the effective rank.   
\medskip
\begin{lemma}
    Let $\xi\geq 2\Vert Z\Vert_2$, we have $k_\xi  \leq n$. If, in addition, $\xi \le {2\over 3}s_{n}(H_\tau)$, then $k_\xi=n$.
    \label{lem:rank}
\end{lemma}

\section{Learning from a single trajectory}\label{sec:single}

In this section, we assume that to estimate the order of the system and its Markov parameters, we have access to the input-ouput sequence corresponding to a single trajectory of the system of length $T$. 

\subsection{Initial Hankel matrix estimation}

We describe here the first step of our procedure that consists in deriving from the data an initial estimator $\hat{H}_\tau$ of $H_\tau$. To this aim, we leverage the framework used in \cite{sarkar2021finite}. Define $H_{t,\tau_1,\tau_2}$ the Hankel matrix of size $(\tau_1d_u, \tau_2 d_y)$ such that its $(i,j)$ block is equal to $CA^{t+i+j-2}B$ for $i=1,...,\tau_1$ and $j=1,...,\tau_2$. Note that we have $H_\tau=H_{0,\tau,\tau}$. Similarly, define the strict lower triangular block Toeplitz matrix 
$T_{t,\tau}$ such that $(i,j)$ block is equal to $CA^{t+i-j-1}B$ if $i>j$ and $0$ otherwise.
Using this notation, we can write the input-output relationship as follows: $\forall t =\tau+1,...,T-\tau+1$,
\begin{align}
& \tilde{y}_t =  H_{0,\tau,\tau}\tilde{u}_t + T_{0,\tau} \begin{pmatrix}
    u_t^\top & 
    \cdots  & 
    u_{t+\tau-1}^\top \end{pmatrix}^\top
\nonumber\\ & + H_{\tau,\tau,t-\tau-1} \begin{pmatrix}
    u_{t-\tau-1}^\top & 
    \cdots & 
    u_1^\top\end{pmatrix}^\top
+ \begin{pmatrix}
    z_t^\top & 
    \cdots & 
    z_{t+\tau-1}^\top
\end{pmatrix}^\top\label{eq:modH}
\end{align}
where $\tilde u_{t} = \begin{pmatrix}u_{t-1}^\top  &  \cdots & u_{t-\tau}^\top\end{pmatrix}^\top \in\mathbb{R}^{\tau d_u}$ and $\tilde{y}_t=\begin{pmatrix}y_t^\top &  \cdots  & y_{t+\tau-1} ^\top \end{pmatrix}^\top \in\mathbb{R}^{\tau d_y}$.
This linear relationship suggests that an initial estimator of $H_\tau=H_{0,\tau,\tau}$ can be obtained using a Least-Squares procedure: 
\begin{equation}
    \hat{H}_\tau = \argmin_{M\in\mathbb{R}^{\tau d_y\times \tau d_u}} \sum_{t=\tau+1}^{T-\tau+1}\Vert \tilde y_t-M\tilde u_t\Vert_{l_2}^2
    \label{eq:lse_h}
\end{equation}
\noindent
As it turns out, the accuracy of $\hat{H}_\tau$ can be analyzed. The following proposition is a corollary of results in \cite{sarkar2021finite} (in our case, we do not have process noise, but have general variances for the inputs and observation noise $\sigma_u,\sigma_z$).

\medskip
\begin{assumption}\label{ass:beta} Let $\mathcal{G}(s)=\sum_{k=1}^\infty CA^{k-1}Bs^{-k}$ be the transfer function of the system. $\Vert \mathcal{G} \Vert_{H_\infty}=\sup_w \Vert\mathcal{G}(jw)\Vert_2$ is upper bounded by $\beta$. 
\end{assumption}

\medskip
\begin{proposition}(Theorem 5.1 in \cite{sarkar2021finite})\label{prop:lse} Let $\delta \in (0,1)$. Under Assumption \ref{ass:beta}, with probability at least $1-\delta$,
$$\Vert \hat H_\tau-H_\tau\Vert_2\leq 4\frac{\max(\beta\sqrt{\tau},\sigma_z )}{\sigma_u}\sqrt{\frac{d_y\tau+d_u+\log(\frac{1}{\delta})}{T}}$$ 
whenever $T\gtrsim T_0:=\tau \log^2(\tau)(d_u^2\log^2(\frac{d_u^2}{\delta}) + \log(\tau))$.
\end{proposition}


\subsection{Thresholded Ho-Kalman algorithm}

We can exploit the result of Proposition \ref{prop:lse} to design the threshold $\xi$ to be applied to the initial estimator $\hat{H}_\tau$. We assume here that the upper bound $\beta$ on the $H_\infty$-norm of the transfer function. Having access to this upper bound essentially means that the practitioner has control over the worst case gain of the system. This assumption is also made by the authors of \cite{sarkar2021finite} when they wish to select the horizon $\tau$ (a model selection problem). Define the threshold:
\begin{equation}\label{eq:thresh}
\xi= 8\frac{\max(\beta\sqrt{\tau},\sigma_z )}{\sigma_u} \sqrt{\frac{d_y\tau+d_u+\log(\frac{1}{\delta})}{T}}.
\end{equation}

\medskip
\begin{theorem}
Suppose that Assumption \ref{ass:beta} holds. Let $\delta\in (0,1)$. Let $\hat H_\tau$ be the LSE defined in (\ref{eq:lse_h}) and set the threshold $\xi$ as in (\ref{eq:thresh}). \\
{\it (a) Hankel estimation error.} If $T\gtrsim T_0$ (as defined in Proposition \ref{prop:lse}), then with probability at least $1-\delta$, 
$$\Vert\hat H_\tau(\xi)-H_\tau\Vert_F^2\lesssim \min_{k=0,...,n} \left(k\xi^2 + \sum_{i=k+1}^{n}s_i^2(H_\tau)\right).$$
{\it (b) Order estimation error.} If $T\gtrsim \max(T_0, T_1)$ where $T_1:= \frac{\max(\beta^2\tau,\sigma_z^2 )}{\sigma_u^2}\frac{d_y\tau+d_u+\log(\frac{1}{\delta})}{s_n^2(H_\tau)}$, then with probability at least $1-\delta$, $\textrm{rank}(\hat{H}_\tau(\xi))=n$.\\
{\it (c) Markov parameter estimation error.} If $T\gtrsim \max(T_0, T_2)$ where $T_2:=\frac{\max(\beta^2\tau,\sigma_z^2 )}{\sigma_u^2}\frac{d_y\tau+d_u+\log(\frac{1}{\delta})}{s_n^2(H_\tau^\rightarrow)}$, then with probability at least $1-\delta$, there exists a unitary matrix $P$ such that the outputs $\hat A$, $\hat B$, $\hat C$ of the thresholded Ho-Kalman algorithm satisfy:
    \begin{align}\label{eq:ho}
    \begin{split}
    \Vert \bar A -P^\top \hat A P\Vert_F & \leq\frac{50 \Vert H_\tau\Vert_2\Vert \hat{H}_\tau(\xi) - H_\tau\Vert_F}{s_n^{2}(H_\tau^\rightarrow)},\\
    \max(\Vert \bar B-P^\top \hat B\Vert_F,\Vert \bar C-\hat CP\Vert_F) &\leq \frac{\sqrt{5}\Vert \hat{H}_\tau(\xi) - H_\tau\Vert_F}{\sqrt{s_n(H_\tau^\rightarrow)}},
    \end{split}
\end{align}
where $\bar{A}$, $\bar{B}$, $\bar{C}$ are the outputs of the Ho-Kalman algorithm applied to the true Hankel matrix $H_\tau$. 
     
    \label{thm:sing_traj}
\end{theorem}

\medskip
The statements (a) and (b) in Theorem \ref{thm:sing_traj} provide upper bounds of the sample complexity for the estimation of the Hankel matrix and the order of the system. When it comes to the estimation of the Markov parameters, the authors of \cite{oymak2019non} derive error bounds when the order of the system $n$ is initially known (in this case, the Ho-Kalman algorithm is applied to $\Pi_n(\hat{H}_\tau)$). The statement (c) of our theorem establishes that our thresholded Ho-Kalman algorithm yields the same error upper bounds without the knowledge of the system order.

\section{Learning from multiple trajectories}\label{sec:multi}

Next, we study the case where the data available consists in multiple trajectories. The system identification problem in this setting has been extensively investigated, e.g., in \cite{tu2017non,zheng2020non,sun2022finite}.

\subsection{Initial Hankel matrix estimation}

To estimate the system parameters, we have access to $T'=\lfloor \frac {T}{2\tau-1}\rfloor$ trajectories, each of length $2\tau-1$. This ensures a fair comparison with the single trajectory setup as the total number of samples available is $T$. To simplify the notation and analysis, we assume that each trajectory starts with state $x_1=0$. Under this assumption, the dynamics can be completely unrolled so that the output can be written based on the inputs only. Indeed, by taking $t=2\tau$ in (\ref{eq:G}), we get: 

$$
y_{2\tau}=G_\tau \begin{pmatrix}
    u_{2\tau-1}^\top &  
    \cdots & 
    u_{1}^\top 
\end{pmatrix}^\top  + z_{2\tau}. 
$$ 

Again this linear relationship suggests a Least-Squares procedure to estimate $G_\tau$. We can then apply the operator $\mathcal{H}$ to the resulting estimator to derive $\hat{H}_\tau$, as proposed in \cite{sun2022finite} and outlined below. Define $\bar u_{2\tau} := \begin{pmatrix}
    u_{2\tau-1}^\top &  
    \cdots & 
    u_{1}^\top 
\end{pmatrix}^\top\in\mathbb{R}^{(2\tau-1)d_u}$. For each $i=1,...,T'$, we collect $(y_{2\tau}^{(i)},\bar u_{2\tau}^{(i)})$ and compute the LSE
\begin{equation}  \hat{G}_{\tau}=\argmin_{M\in\mathbb{R}^{d_y\times(2\tau-1)d_u}} \sum_{i=1}^{T'}\Vert y_{2\tau}^{(i)} - M\bar u_{2\tau}^{(i)}\Vert_{l_2}^2.
\label{eq:lse_g}
\end{equation}  
One can establish high probability upper bounds of the error of $\hat{H}_\tau:=\mathcal{H}(\hat{G}_{\tau})$. This was initially done in Theorem 3 of \cite{sun2022finite}, albeit with un-precised probability and conservative sample complexity $T'\geq \tau d_u d_y$, which we refine using tight concentration results.


\medskip
\begin{proposition}\label{prop:multi} Let $\delta\in (0,1)$.
With probability at least $1-\delta$,
$$\Vert \hat H_\tau-H_\tau\Vert_2\leq 2\frac{\sigma_z}{\sigma_u}\sqrt{\frac{\min(d_y,\tau) (\tau d_u + \log(\frac{1}{\delta}))}{T'}} $$
whenever $T\gtrsim T_0=\tau(\log(\frac{1}{\delta})+(2\tau-1)d_u)$.
\end{proposition}

\subsection{Thresholded Ho-Kalman algorithm}

As in the previous section, we exploit the concentration result of Proposition \ref{prop:multi} to define the threshold: 

\begin{equation}\label{eq:thresh_mult}
\xi= 4\frac{\sigma_z}{\sigma_u} \sqrt{\tau\frac{\min(d_y,\tau) (\tau d_u + \log(\frac{1}{\delta}))}{T}}.
\end{equation}

\begin{theorem} 
Let $\delta\in (0,1)$. Let $\hat G_\tau$ be the LSE defined in (\ref{eq:lse_g}) and set the threshold $\xi$ as in (\ref{eq:thresh_mult}).\\
{\it (a) Hankel estimation error.} If $T\gtrsim T_0$ (as defined in Proposition \ref{prop:multi}), then with probability at least $1-\delta$, 

$$\Vert\hat H_\tau(\xi)-H_\tau\Vert_F^2\lesssim \min_{k=0,...,n} \left(k\xi^2 + \sum_{i=k+1}^{n}s_i^2(H_\tau)\right).$$
{\it (b) Order estimation error.} If $T\gtrsim \max(T_0, T_1)$ where $T_1:= \frac{\sigma_z^2 }{\sigma_u^2}\frac{\tau\min(d_y,\tau) (\tau d_u + \log(\frac{1}{\delta}))}{s_n^2(H_\tau)}$, then with probability at least $1-\delta$, $\textrm{rank}(\hat{H}_\tau(\xi))=n$.\\
{\it (c) Markov parameter estimation error.} If $T\gtrsim \max(T_0, T_2)$ where $T_2:=\frac{\sigma_z^2 }{\sigma_u^2}\frac{\tau\min(d_y,\tau) (\tau d_u + \log(\frac{1}{\delta}))}{s_n^2(H_\tau^\rightarrow)}$, then with probability at least $1-\delta$, there exists a unitary matrix $P$ such that the outputs $\hat A$, $\hat B$, $\hat C$ of the thresholded Ho-Kalman algorithm satisfy identification error bounds (\ref{eq:ho}).
    \label{thm:mult_traj}
\end{theorem}

\medskip
Theorem \ref{thm:mult_traj} presents results similar to those for system identification from a single trajectory. However, in the case of estimation from multiple trajectories, the sample complexity is slightly worse — this arises from the term $\min(d_y,\tau)$ in the threshold, which is likely conservative. An interesting direction for future work is to determine whether this term can be removed. Note finally that the threshold used in the multiple trajectory setting does not require the knowledge of $\beta$ (an upper bound on the $H_\infty$ norm of the transfer function). 
\section{Numerical experiments}

In this section, we illustrate the performance of our thresholded Ho-Kalman algorithm in the multiple trajectory setting. For the sake of simplicity, we consider a synthetic example inspired from \cite{oymak2019non} where $n=5$, $d_y=2, d_u=3$. Values of $\sigma_u=1$, $\sigma_z=0.1$, $\tau=6$ are fixed. $A$ is chosen diagonal with entries independently drawn uniformly at random between $0.1$ and $0.9$, which ensures stability of the system. Entries of $B$ and $C$ are drawn independently from a normal distribution with zero mean and standard deviation equal to $2$. 


One {\it trial} consists in exploiting $T$ trajectories of size $2\tau-1$ to compute $\hat{H}_\tau(\xi)$. One {\it experiment} consists in $R=20$ trials. We repeat the experiment for various values of $T$ and record the outputs, i.e., the effective rank $k_\xi$ and the estimators of the Markov parameters.



\begin{figure}
    \centering
    \includegraphics[width=0.5\linewidth]{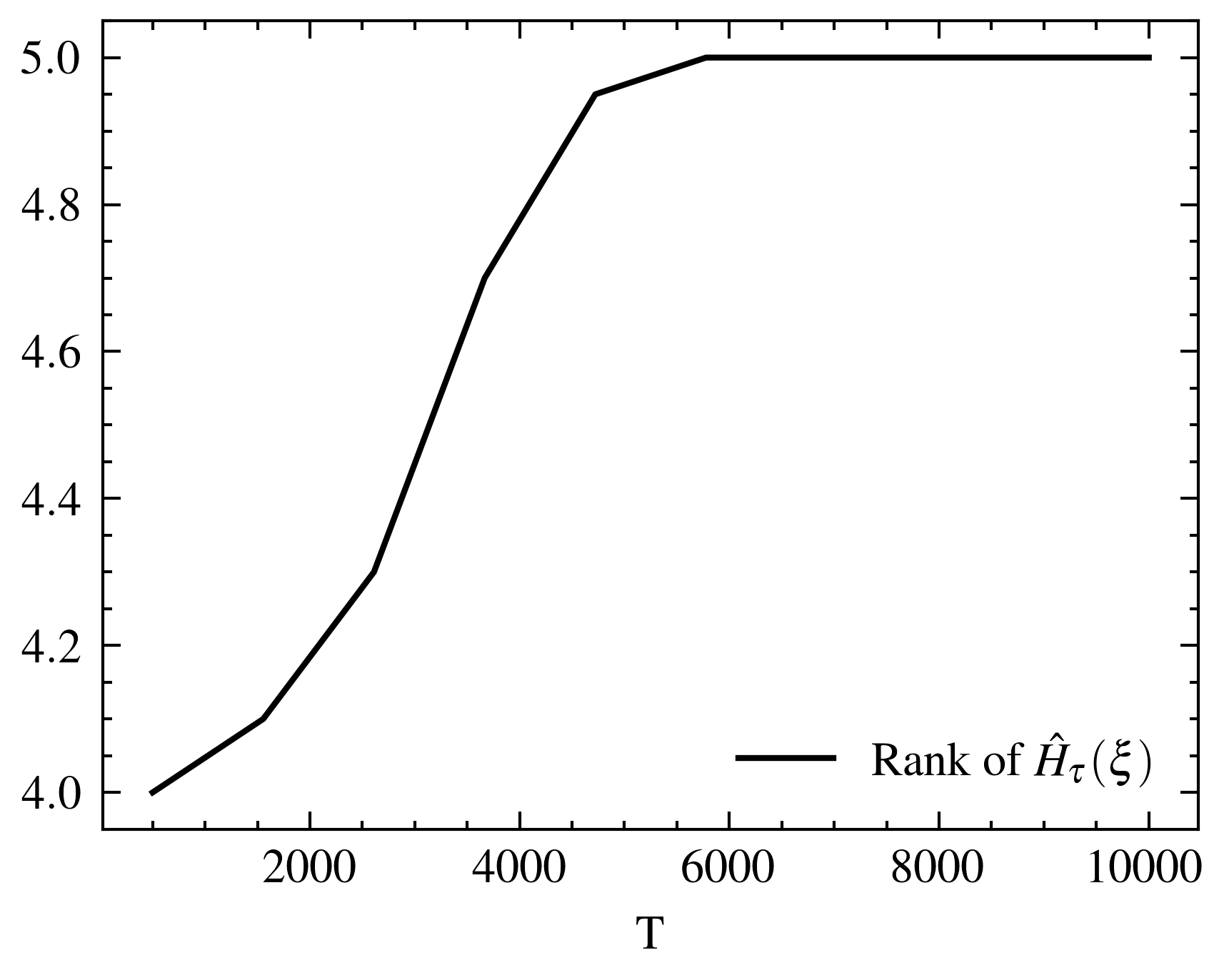}
    \vspace*{-3mm}
    \caption{Order recovery vs. sample size $T$}
    \label{fig:k_xi}
\end{figure}
In Figure \ref{fig:k_xi}, we plot the evolution of the estimated order $k_\xi$ when $T$ increases (and therefore $\xi$ decreases). Since, it is an average over several trials, the value is not necessarily an integer. As expected, we observe that $k_\xi$ converges towards thre true order $n$ and is exactly equal to the latter after $\sim 5000$ samples.

\begin{figure}
    \centering
    \includegraphics[width=0.55\linewidth]{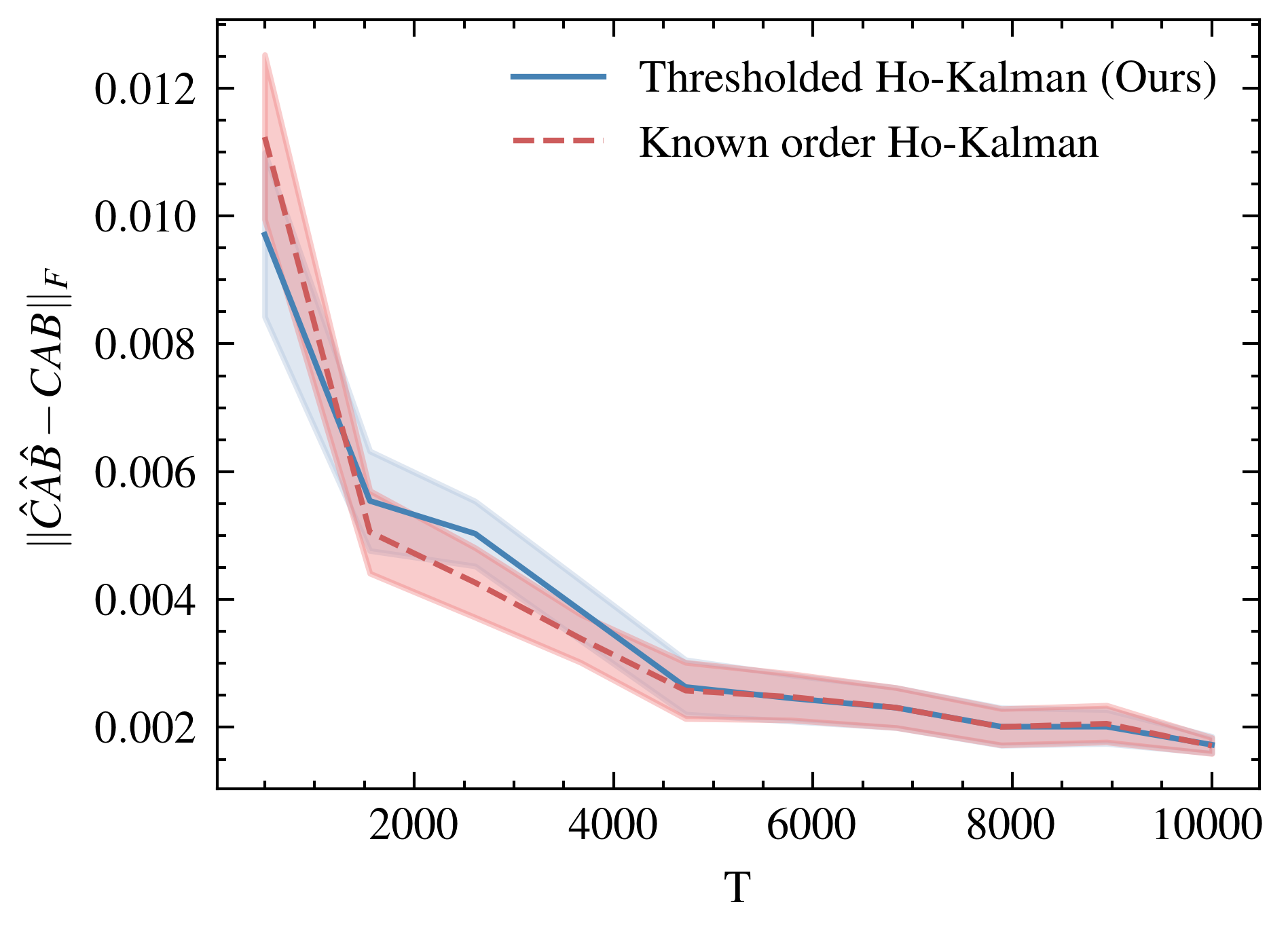}
    \vspace*{-3mm}
    \caption{The performance of the thresholded Ho-Kalman algorithm vs. sample size $T$. Comparison with an Oracle algorithm aware of the order.}
    \label{fig:err_cab}
\end{figure}
When analyzing the estimation of the Markov parameters, it is not possible to compute individual errors for $\hat A$, $\hat B$, $\hat C$ since identification is up to an invertible transformation. Instead, we plot in Figure \ref{fig:err_cab} $\Vert \hat C\hat A \hat B -CAB\Vert_F$. This expression also has the advantage of being well defined even for small values of $T$ where we do not have $k_\xi=n$ yet. We compare our algorithm to that proposed in \cite{oymak2019non} where the order is known and used. As expected, for $T\gtrsim 5000$ samples, both versions have identical performances since our algorithm estimates the order correctly. Surprisingly, for lower values of $T$ where we do not estimate the order accurately, our algorithm still performs almost as well as if the true order was known. 
\section{Conclusion}

In this paper, we introduced the Thresholded Ho-Kalman algorithm, designed to estimate partially observed Linear Time-Invariant (LTI) systems from their input-output trajectories. The algorithm leverages a singular value thresholding procedure to ensure the estimated Hankel matrix has an appropriate effective rank. It also estimates the minimal system order and its Markov parameters, and enjoys finite-time performance guarantees. To assess the statistical optimality of our algorithm, an important future research direction is to establish fundamental limits on the error bounds for estimating the Hankel matrix and its rank. Fundamental limits have been identified for fully observable systems \cite{jedra2022finite}, but the challenge is significantly greater in the partially observed case due to the complex statistical properties of the noise in the input-output relationship (see, e.g., (\ref{eq:modH})).

\bibliographystyle{plain}
\bibliography{biblio}




\appendix


\newpage
\noindent
\section{Proofs}
\subsection{Proof of Proposition \ref{prop:threshold}}
For any $k=1,...,n$, we have by the Eckart-Young's theorem:
$$\Vert \Pi_k(H_{\tau}+ Z) -H_{\tau}-Z\Vert_F^2 \leq  \Vert \Pi_k(H_{\tau}) - H_{\tau} - Z \Vert_F^2.$$ 
By decomposing the squares, we obtain: 
\begin{align*}
    \Vert \Pi_k(H_{\tau}+Z) -H_{\tau} \Vert^2 &\leq \Vert H_{\tau}-\Pi_k(H_{\tau})\Vert_F^2   +2 \langle \Pi_k(H_{\tau}+Z) -\Pi_k(H_{\tau}), Z \rangle.
\end{align*}
For the l.h.s, we have, by reverse triangle inequality,
$$
    \textrm{l.h.s} \geq \left| \Vert \Pi_k(H_{\tau}+Z) -\Pi_k(H_{\tau})\Vert_F -  \Vert H_{\tau} -\Pi_k(H_{\tau})\Vert_F \right|^2.
$$
Let $\varphi= \langle \Pi_k(H_{\tau}+Z) -\Pi_k(H_{\tau}), Z \rangle $. Since $\Pi_k(H_{\tau}+Z) -\Pi_k(H_{\tau})$ is of rank at most $2k$, we have by Lemma 2 of \cite{xiang2012optimal}
$$    \varphi 
\leq \Vert \Pi_k(H_{\tau}+Z) -\Pi_k(H_{\tau})\Vert_F \Vert \Pi_{2k}(Z)\Vert_F.
$$
Plugging these bounds together and decomposing the square, we obtain: 
$$        \frac{1}{2}\Vert \Pi_k(H_{\tau}+Z) -\Pi_k(H_{\tau})\Vert_F \leq \Vert \Pi_{2k}(Z)\Vert_F+ \Vert H_{\tau} -\Pi_k(H_{\tau})\Vert_F.
$$
We conclude by noting that:
\begin{align*}
\Vert \Pi_k(\hat H_\tau)-H_{\tau}\Vert_F &\leq \Vert \Pi_k(H_{\tau}+Z) -\Pi_k(H_{\tau})\Vert_F \\
&+ \Vert H_{\tau} -\Pi_k(H_{\tau})\Vert_F \\
&\leq 2\Vert \Pi_{2k}(Z)\Vert_F + 3\Vert H_{\tau} -\Pi_k(H_{\tau})\Vert_F.
\end{align*}
Let $\alpha\geq 0$ and $K=\max\{i:s_i(\hat H_\tau)\geq (2+\alpha)\Vert Z\Vert_2\}$. Define $k^{\star}=\max\{i:s_i(H_{\tau})\geq (3+\alpha)\Vert Z\Vert_2\}$. By Weyl's inequality, we have:

\begin{itemize}
    \item $s_{K+1}(H_{\tau})\leq s_{K+1}(\hat{H}_{\tau}) + \Vert Z\Vert_2\leq (3+\alpha)\Vert Z \Vert_2$, which implies that $k^{\star}\leq K$; 
    \item $\forall i\leq K, \ s_i(H_{\tau})\geq s_i(\hat H_\tau)-\Vert Z\Vert_2\geq (1+\alpha)\Vert Z\Vert_2 $.
\end{itemize}  
Hence, we can write:
\begin{align*}
    \frac{\Vert \Pi_{K}(\hat H_{\tau})-H_{\tau}\Vert_F^2}{18}
    &\leq K\Vert Z\Vert_2^2+ \sum_{i>K}s_i^2(H_{\tau}) \\ 
    &= k^{\star}\Vert Z\Vert_2^2+\sum_{i>k^{\star}}s_i^2(H_{\tau}) +\psi\\
    &\leq k^{\star}\Vert Z\Vert_2^2+\sum_{i>k^{\star}}s_i^2(H_{\tau}) \\
    &\leq (3+\alpha)^2k^{\star}\Vert Z\Vert_2^2+\sum_{i>k^{\star}}s_i^2(H_{\tau}) \\
    &= \min_{k=0,...,n} \left\{(3+\alpha)^2k\Vert Z\Vert_2^2+ \sum_{i=k+1}^ns_i^2(H_{\tau})\right\} 
\end{align*}
where $\psi=\sum_{i=k^{\star}+1}^{K}\left(\Vert Z\Vert_2^2-s_i^2(H_{\tau})\right)<0$ by the second point above. Let now $\xi\geq 2\Vert Z\Vert_2$ and $\alpha=\frac{\xi}{\Vert Z\Vert_2}-2$ then $\xi =(2+\alpha)\Vert Z\Vert_2$ and we apply the above by observing that $(3+\alpha)\Vert Z\Vert_2=(1+\frac{\xi}{\Vert Z\Vert_2})\Vert Z\Vert_2\leq 2\xi$.
\hfill{$\Box$}

\medskip
\noindent
\subsection{Proof of Lemma \ref{lem:rank}}
By Weyl's inequality, we have $s_{n+1}(\hat H_\tau)\leq s_{n+1}(H_\tau)+\Vert Z\Vert_2 = \Vert Z\Vert_2\leq \frac{\xi}{2}$. Therefore $k_\xi< n+1$. If $s_{n}(H_\tau)\geq \frac{3}{2}\xi$ then $s_{n}(\hat H_\tau)\geq s_{n}(H_\tau)-\Vert Z\Vert_2\geq \frac{3}{2}\xi-\frac{\xi}{2}=\xi $. Therefore $k_\xi\geq n$.\hfill{$\Box$}


    



\medskip
\noindent
\subsection{Proof of Theorem \ref{thm:sing_traj}}
    The first two points are straightforward applications of Proposition \ref{prop:lse} combined with Proposition \ref{prop:threshold} and Lemma \ref{lem:rank}. We focus on the third point.
\\    
The strategy is identical to the one adopted in \cite{oymak2019non}, except that we  need to additionally show $\hat H_\tau(\xi)^\rightarrow$ has rank $n$ and verifies $\Vert \hat H_\tau(\xi)^\rightarrow-H_\tau^\rightarrow \Vert_2 \leq \frac{s_n(H_\tau^\rightarrow)}{2}$. 
\\  
Since $s_n(H_\tau^\rightarrow )\leq s_n(H_\tau)$ as the former is a submatrix of the latter, then $T_2\geq T_1$, and by Theorem \ref{thm:sing_traj}, we have with probability at least $1-\delta$:
    $$\xi\geq 2\Vert \hat H_\tau-H_\tau\Vert_2 \qquad \textrm{and} \qquad \mathrm{rank}(\hat H_\tau(\xi))=n$$  
On the event $\{\mathrm{rank}(\hat H_\tau(\xi))=n\}$, we have: 
    \begin{align*}
        \Vert \hat H_\tau(\xi)^\rightarrow-H_\tau^\rightarrow \Vert_2 &\leq \Vert \hat H_\tau(\xi)-H_\tau \Vert_2 \\
        &\leq \Vert \hat H_\tau(\xi)-\hat H_\tau \Vert_2 + \Vert \hat H_\tau(\xi)-H_\tau \Vert_2 \\
        &\leq 2\Vert \hat H_\tau-H_\tau \Vert_2 \leq 4\xi \leq \frac{1}{2}s_n(H_\tau^\rightarrow),
    \end{align*}
where the first inequality holds as $H_\tau(\xi)^\rightarrow-H_\tau^\rightarrow$ is a submatrix of $H_\tau(\xi)-H_\tau$, the third holds by Eckart-Young's Theorem since $\hat H_\tau(\xi)$ is the best rank $n$ approximation of $\hat H_\tau$, and the last one by definition of $T_2$. By Weyl's inequality, 
\begin{align*}
    s_n(\hat H_\tau(\xi)^\rightarrow)&\geq s_n(H_\tau^\rightarrow)-\Vert \hat H_\tau(\xi)^\rightarrow-H_\tau^\rightarrow \Vert_2  \geq \frac{1}{2}s_n(H_\tau^\rightarrow) >0.
\end{align*}
Hence $\mathrm{rank}(\hat H_\tau(\xi)^\rightarrow)\geq n$ and since $\mathrm{rank}(\hat H_\tau(\xi))= n$ then we actually have equality in the inequality.
The remaining steps of the proof follow mutatis mutandis the proof of Theorem 5.3 and Corollary 5.4 in \cite{oymak2019non}. We  restate the main lines for sake of brevity. Let $U,S,V, O, Q$ be the oracle counterparts obtained from the SVD of $H_\tau^\rightarrow$ . By Lemma 5.14 of \cite{tu2016low}, there exists a unitary  $P\in\mathbb{R}^{n\times n}$ such that 
$$
    \Vert US^{1/2}-\hat U \hat S^{1/2}P\Vert_F^2 + \Vert VS^{1/2}-\hat V\hat S^{1/2}P\Vert_F^2 
    \leq\frac{5\Vert \hat H_\tau(\xi)^\rightarrow - H_\tau ^\rightarrow\Vert_F^2 }{s_n(H_\tau^\rightarrow)}
$$
We directly deduce the corresponding bounds on $\hat C$, $\hat B$ since these are submatrices of $\hat U \hat S^{1/2}$ and $\hat S^{1/2}\hat V^\top$. The error for $\hat A$ also hinges on this result but is slightly more technical, we refer to \cite{oymak2019non} for exact computations.
\hfill{$\Box$}


\medskip
\noindent
\subsection{Proof of Proposition \ref{prop:multi}}
    We relate the LSE error $Z=\hat G_\tau-G_\tau$ to the Hankel error. If $\tau\leq d_y$ then we use the following inequality, Lemma 5.2 of \cite{oymak2019non}, $\Vert \hat{H}_\tau-H_\tau\Vert_2\leq \sqrt{\tau} \Vert Z\Vert_2$. If $d_y\leq \tau$ then we use Theorem 9 of \cite{sun2022finite}: Let $F$ be the unitary discrete Fourier matrix then $\Vert \hat H_\tau-H_\tau\Vert_2\leq \Vert \begin{bmatrix}\Vert FZ_1 \Vert_\infty & \cdots & \Vert FZ_{d_y}\Vert_\infty  \end{bmatrix} \Vert_2$ where $Z_i$ corresponds to the $i$-th column of $Z$. In particular, since $F$ is unitary and the $Z_i$ have identical distribution, this gives 
    $\Vert \hat H_\tau-H_\tau\Vert_2\leq \sqrt{d_y}\Vert Z_1 \Vert_2$. By Lemma 3 in \cite{bunea2011optimal}, we have with probability at least $1-\delta$ that $\Vert Z_1\Vert_2\leq 2\sigma_z\sqrt{\frac{\tau d_u + \log(\frac{1}{\delta})}{T'\lambda_\mathrm{min}(\hat \Sigma)}}$ where $\hat{\Sigma}=\frac{1}{T'}\sum_{i=1}^{T'} u_{2\tau}^{(i)}(u_{2\tau}^{(i)})^\top$. Since the inputs are assumed i.i.d Gaussian then each $u_{2\tau}^{(i)}\sim N(0,\sigma_u^2I_{(2\tau-1)d_u})$, and assuming $T\geq 4(2\tau-1)d_u+4\log(\frac{1}{\delta})$, we use Theorem 8 in \cite{barzilai2024simple} to replace $\lambda_\mathrm{min}(\hat \Sigma)$ by $\sigma_u^2$.
\hfill{$\Box$} 

\medskip
\noindent
\subsection{Proof of Theorem \ref{thm:mult_traj}} The proof follows the same lines as that of Theorem \ref{thm:sing_traj}.\hfill{$\Box$}



\end{document}